# Toward Ethical AIED

Conclusions to *The Ethics of Artificial Intelligence in Education: Practices, Challenges and Debates*

*(Wayne Holmes and Kaśka Porayska-Pomsta Eds., August 2022, Routlege)*

by
Kaśka Porayska-Pomsta and Wayne Holmes

University College London, IOE UCL's Faculty of Education and Society, UCL Knowledge Lab

k.porayska-pomsta@ucl.ac.uk; w.holmes@ucl.ac.uk

In 1960, Norbert Weiner discussed "Some Moral and Technical Consequences of Automation" (Weiner, 1960). A pioneer of cybernetics, often credited with an early formulation of intelligent behaviour (of living things and machines) as a feedback mechanism, Weiner made two related observations which are not only pertinent to this day, but whose implications have since acquired a somewhat prophetic quality.

The first observation relates to learning and adaptive capabilities as pre-requisites of human-like 'intelligent' computers whereby, to be considered artificially intelligent, a machine must be able to learn from and to act on the environment in ways that maximise the probability of it achieving some pre-specified objective. This first point has been well rehearsed and refined over the past 60 years (see e.g., Legg and Hutter, 2007; Russell and Norvig, 1995; Russell, 2019) in what has subsequently become known as the field of Artificial Intelligence. The outcomes and implications of these rehearsals continue to emerge in diverse forms of AI innovations and applications, including in the context that is central to this book, namely AI in education. One implication, for better or worse, that relates to Weiner's first observation, is the ever growing 'smartness' of technology, which often transcends some human capabilities, *nota bene* the technology's capacity to track and harness at speed a growing number of pasts in some specific contexts to predict many possible futures in those same contexts (e.g., Russell, 2019).

The significance of Weiner's second observation, which links directly to the implication stated above, remained largely obscured to AI practitioners until the recent advances in AI (*sic* machine learning) and the shift of AI from research laboratories into mainstream usage. This second point concerns the idea that as machine capabilities grow and as technologies become smarter, our ability to keep pace with and to understand their operations decreases. This gap between the increasing machine smartness and our ability to keep up with it is time constant in the sense that while machines' performance accelerates both in speed and accuracy on specific tasks, human abilities remain comparatively unchanged and slow. Put together, the two observations highlight a real possibility of an event horizon for humans (a singularity), whereby in Weiner's own words: "by the time we are able to react to information conveyed by our senses and stop the car we are driving, it may already have run head on into a wall." (Weiner, 1960, p. 81). Wiener's profound and to date largely disregarded take-home caution here is that:

> "***If we use, to achieve our purposes, a mechanical agency with whose operation we cannot interfere once we have started it***, because the action is so fast and irrevocable that we have not the data to intervene before the action is complete, ***then we had better be quite sure that the purpose put into the machine is the purpose which we really desire*** and not merely a colourful imitation of it." (Weiner, 1960, p.88; authors' emphasis).

While it might be tempting to interpret Weiner's caution as a doomsday exaggeration, given the increasingly well documented concerns about AI's operation and the impact thereof in diverse contexts, it is difficult to ignore the palpability of the situation it refers to. At a general level, there are multiple questions which arise from this caution. One such question is philosophical in nature, as it relates to what we consider the desired purpose (or purposes) to be. To address this question demands a continuous (re-)interrogation of our values from different epistemological perspectives (moral, social, economic, individual, collective, etc.) and at different levels of granularity – from high level aspirations for the society down to low-level tangible outcomes for individuals. As has been discussed throughout this book, such questioning is non-trivial for the answers will depend on who is formulating and addressing the questions, when and why (see Chapter 6 by Holstein and Doroudi). Such questioning is also likely impossible to lead to conclusive answers, or to solutions that simultaneously or permanently satisfy all concerns and all concerned (see Chapter 7 by Kizilec and Hansol).

Another important question that arises from Weiner's reflections is of a civil engineering nature, as it interrogates what AI technologies we design and how we ensure that they are fit for our purpose (assuming that the purpose is known). Indeed, the question relates to whether and how we build AI tools to address human challenges in ways that are beneficial not merely by intention, but by their design, operation and, critically – by the feedback that comes from their actual usage.

In his recent work, Stuart Russell returned to Wiener's observations, both to debunk the popular notion of AI as a human mirroring artefact, and to question the fundamental assumptions of the field as a whole (see for example Russell, 2019; ATI lecture[1]). In doing so, Russell draws a precise link between those fundamental assumptions (which he declares mistaken) and the moral and ethical challenges related to AI technologies. Specifically, he emphasises and reiterates Weiner's second observation by stating that while AI already exceeds some human capabilities, and at some point in the future it may exceed them all (although currently there is scant evidence of that, Vallor, 2021), there will never be a time when human and AI capabilities will be comparable. Where AI is growing computationally powerful, humans remain computationally limited; where AI remains rigid in its preferences, human preferences have high plasticity; where human preferences and inferences are hierarchically structured, those of AI are often not, and so on.

---

[1] https://www.youtube.com/watch?v=_H87qqT8pdY&t=1s

To try to compare AI and humans is not only misguided, it also fuels an unwanted division between AI, the ethics of AI and the ethics of the broader socio-economic systems (including the educational system) within which AI operates. This is because, such comparison diverts us from the fact that much human activity, including the technological inventions that are used to aid or enhance such activity, is based on the same standard model – namely one that involves setting an objective and devising means which single-mindedly maximise the achievement of that objective. The standard model we use in AI is rooted deeply within the economics of our existence, where success is evaluated against discrete and measurable benefits and utilitarian values (typically using economic values of cost, loss, utility etc.) that are fixed into the way we define and go about achieving our objectives in society more generally. As Russell explains, the issue with this standard model being applied in AI lies in its rigid insistence on solving problems within local optima of the limited contexts within which AI operates. It also lies in human fallibility in specifying the objectives that can be guaranteed to be ethical and beneficial to all possible stakeholders at all times, or that can extend towards pursuing global optima that are native to the complex nature of human environments and activities (see also Chapter 1 by Treviranus).

There are many examples in AI, and indeed in other domains, which demonstrate these issues. In AI, this has become manifest through some more obvious examples of the choices of problems with which AI has been tasked, that are ethically questionable at their base, such as the infamous example of using face recognition to predict criminality. Some less obvious examples include defining the objective function to maximise click-throughs, e.g., on social media. Far from having innocent consequences, this has led to a marked proliferation of divisions between people and an entrenchment of people's prejudices, by the sheer fact that by feeding people content that they are likely to click on, their pre-existent biases tend to be reinforced. Here, Russell draws our attention to how the standard model of AI lies at the core of even the simplest machine learning algorithms, making people more uniform and predictable; instead of merely pushing content that the users want, the machine learning algorithms that underpin these kinds of applications fundamentally modify people's views towards more predictable and eventually towards more extreme positions, hence creating a much easier environment for AI to classify and to achieve its objectives.

Given that within the standard model, the sole true objective of AI is to optimise the local environment to maximise the chances of success therein, it stands to reason that, if the objective we assign to the AI is wrong or harmful to us (or some of us) by some definition, then the smarter the AI, the worse the outcomes will be for us. Linking back to Weiner's car crash metaphor, given human neuro-cognitive make-up that predisposes us to act on first impressions and automating habitual processes such as the way we interpret our environment and make decisions (see e.g., Houdé et al., 2000), we may not realise in time how we are being changed by this standard optimisation model being enacted on us. In this context, one blind-spot in how we study the benefits and pitfalls of our relationship with AI seems to lie in our lack of understanding of and available research on how AI interacts with human psychology – a point alluded to by Holstein and Doroudi in Chapter 6.

As is discussed and examined throughout this book, AI in Education is not immune to the concerns raised by Weiner and Russell. Furthermore, it represents one of the very high-stakes areas in which these concerns must be examined and addressed urgently, given that education not only shapes life-long thinking and action of individuals from a young age, but it is also an obligatory element of every person's development, and a fundamental human right. Jutta Treviranus' eloquent discussion in Chapter 1 speaks of the overarching ethical concerns for AI in education deriving from its unexceptional position within the broader AI domain insofar as its dominant investment in developing tools such as tutoring systems which optimise "the path to the dead end of the local optima". Such local optima are defined by the demands of the pre-existent socio-economic contexts, of which education forms an integral part. Her commentary seems not so much about tutoring systems being wrong or unethical in themselves; rather it is about the lack of recognition (mainly outside of the AIED community) that they are limited in terms of their pedagogies, their target domains and target users.

## AIED: a solution and/or applied philosophy?

Intelligent tutoring systems represent a *low hanging fruit* for the EdTech industry and business-driven educational policies, as they easily fit into the established interpretation of what learning is in terms of drill and practice mastery learning towards exams. Intelligent tutoring systems do not challenge the status quo of the educational system, and thus, their offering is relatively easily monetised. And yet, as is clear from du Boulay's defence (Chapter 9) of the AIED pedagogies and his discussion of the related ethical dimensions, the pedagogies employed in intelligent tutoring systems represent only one example researched and designed for in the field. Other forms of pedagogies include exploratory learning, collaborative learning, enquiry learning, learning by teaching, etc., (see also the Introduction to this volume by Holmes and Porayska-Pomsta), each involving diverse and often nuanced pedagogical strategies and tactics, and many recruiting tools such as open learner models (e.g., Bull and Kay, 2016), learning by teaching (Biswas et al., 2005), or nuanced help-seeking approaches (Aleven et al., 2016) that aim to foster in students critical thinking through reflection, self-monitoring and self-regulation – albeit that relatively few of these have graduated from the lab to become commercial or widely available tools. Thus, as du Boulay explains, the last twenty years have seen a growing diversification of AIED pedagogies, with AIED researchers investing greatly both in understanding what best pedagogical practices are in different contexts and in trying to define and support what Seymour Pappert called the *art of learning*, which involves active construction of knowledge, discovery, learning from mistakes, and metacognitive competencies. This investment stands in contrast with many commercial EdTech practices and claims, which as Paulo Blinkstein highlighted in his invited talk at the International Conference on Artificial Intelligence in Education in 2018, reflect the industry's push to seize on the seductive allure of AI as a commercial highlight to deliver often half-baked and educationally questionable quack 'solutions'. In his chapter, du Boulay channels growing calls for the need to establish auditing processes for AIED to monitor the quality of the educational offerings from the EdTech industry and AIED research community.

One question that arises from du Boulay's discussion relates to what should be the role of AIED research, if any, in establishing best AIED practices and related ethics auditing processes? How should the AIED research community position its contribution with respect to the growing appetite of a money-making industry and stop-gap fixes by governments to introduce evidence- and knowledge-poor, self-proclaimed miracle cures to the challenges and ailments of the educational systems? These questions are not intended to dismiss the potential of the EdTech industry or AI-related education policies. Instead, in line with Blinkstein's reflections, they aim to help us pause and consider what we want and need respectively from policy, from the EdTech, and from the AIED research community. While policy tends to be an enabling force for change in the real-world contexts, and while EdTech may have better financial resources to deliver at speed AIED applications that work in the real-world, AIED research allows us to consider fundamental questions about how humans learn and develop, how we can support learning and development of students across different ages and in different domains, what constitute best pedagogical practices and contexts that are conducive to learning, how AI tools might be designed specifically to support teachers rather than to undertake teacher tasks, and so on (for example of this see Aleven et al., 2016, cited earlier). In asking such questions, AIED's identity emerges not merely as a design science where interventions are being developed to improve educational outcomes, but as a form of applied philosophy where challenges are being identified and positioned in wider human contexts, and where relevant theories are being developed.

The view of AIED as an applied philosophy aligns with Smuha's discussion (Chapter 5) and with Treviranus' suggestion in Chapter 1 that the way out of the dead end of the local optima is to diversify our perspectives, to be willing to fail, to change paths and strategies to collaborate across differences, and to find the courage and means to extend our thinking beyond the categorical, so that we can embrace human difference as a strength and as a welcome catalyst for innovation and cultural progress, rather than as a hindrance (see also Porayska-Pomsta and Rajendran, 2019; and Mau, 2019). Treviranus' manifesto for the future of AI in Education involves key questions such as what do we want to automate, accelerate and optimise in education, what we are willing to remove from the current practices, and how the purposes we put into the teaching machines will ultimately shape who we are and how we function individually and collectively. Through this she draws attention to the fact that the way we design and deploy AI in education and what ethical questions we interrogate in this context are not separate from the questions about the values and moral drivers that we must ask of the educational system itself in order to then inform how such systems might be served by what she refers to as the AI power tools.

The call for change in the way that we think about the role and designs of AI in Education systems in either amplifying or alleviating broader socio-technical and economic inequalities and exclusive practices is a running theme throughout this book. Each chapter offers an important contribution towards developing a clear, transdisciplinary understanding of what it is that AI does and may still contribute to the context of supporting human learning and development. Crucially, each chapter rehearses distinct questions about AI in Education within the broader socio-economic context, spotlighting key issues from different standpoints. From these different perspectives, we can begin to piece together an initial sketch of AI in Education's strengths and weaknesses, considered common across the different perspectives, and we can identify the blind spots for the field with respect to ethics.

In Chapter 6, Holstein and Doroudi provide a detailed map of the stakeholder perspectives whose voices are crucial to our gaining a better view on the questions we may need to ask of AI in Education as operating in much larger socio-technical settings. In this, they implicitly highlight the diverse nature of the pronoun 'we' that is routinely used in calls to action (also prevalent in this book) in AI in Education. Their discussion and the roadmap towards fairer AIED highlight the need for transparency and precision in declaring whose perspective is being emphasised, in pointing out that each perspective employs different methods that may lead to very different conclusions. Treviranus (Chapter 1), Brossi, Castillo and Cortesi (Chapter 4), and Madaio, Blodgett, Mayfield and Dixon-Román (Chapter 8), each apply different magnifying glasses to explicate that too often the perspectives and the diverse needs of the key intended beneficiaries, namely learners, are invisible to the designers of curricula, and by extension – to the designers of AI systems that implement those curricula. Brossi and colleagues focus on the need for the active participation of young people in the design, deployment, and evaluation of AI systems which they are asked to use, but they point that the lack of such participation reflects established 'adults-know-best' assumptions within the pre-existent systems that all too often lead to young people being disenfranchised from their own educational experiences. Madaio et al., talk about frequent erasure of learners from minority groups by the sheer fact that mainstream curricula are designed typically by dominant groups for the dominant populations (typically white middle classes), which often hinders the cultural accessibility of the established curricula and related pedagogies to learners from minority groups. Treviranus delivers another example of the exclusive nature of the mainstream educational system, which *de facto* marginalises neurodiverse learners, because it is in essence designed to cater for neurotypical learners, whereby any marked divergence from the so-called 'norm', is understood as a form of deficit and deviation (leading to, but all too often not being recognised as, harms of ex-nomination). The issues raised within those chapters are not exclusive to AIED, rather they highlight issues that pervade the educational and other domains of the broader socio-economic system. As such, the discussions within this book reveal how technology designs mirror the established system and how technology serves as an amplifier of this system. Thus, while AI may not mirror us, it certainly has proven to be able to offer a clear mirror onto ourselves and our practices.

**AIED: a research methodology**

Many of the examples described in this book document biases that are prevalent in educational contexts. Such biases tend to lead to a multitude of consequences related to both harms of allocation (i.e., the inability of some people to access key resources) and harms of representation (i.e., some people's identities and the associated needs not being represented, or being over-represented to the point of stigma), as also outlined in the Introduction to this volume. Such biases also solidify particular research foci within the AIED field as well as the assumption of cultural and neuro-cognitive uniformity of student populations. As is demonstrated throughout this book, a precise contextualisation of these biases is necessary to allow us to question systematically who the different AIED systems serve and how, how they may serve as enablers to some and disablers to others, who benefits and who loses out and why as a consequence of the particular AIED designs and modes of deployment. In turn, such questioning is necessary to improve, update and de-bias our research practices and methods (Fox, Chapter 2; Madaio et al., Chapter 8), to our being able to diversify AIED's research and design methodologies, to facilitate a greater transparency with respect to the strengths and weaknesses of the tools we build to then inform our policies with respect to AIED (see especially Bartoletti's discussion in Chapter 3), and, as Howley, Mir and Peck explain in Chapter 10, to educate AI practitioners for whom ethical considerations of the AIED systems will be as critical as the programming languages they use to develop such systems.

One important characteristic of AIED that is rarely recognised outside of the immediate AIED research community is that it relies on AI models not just to build intelligent learning environments to deliver educational solutions, but also (and in many respects primarily) to address the fundamental (philosophical, theoretical, and practical) questions raised within the field. This methodological aspect of AIED (*AIED as a methodology*; see also Porayska-Pomsta, 2016) is clearly illustrated through Kizilcec and Hansol's patient examination in Chapter 7 of the assumptions related to fairness and equity in education, and AIED more specifically, as encoded in the algorithms. Through examining different models of fairness, they draw attention to the types of assumptions and methodologies that impact on the quality of data and algorithms in AI applications for education. They use this examination to elaborate on the challenges identified in the broader AI context with respect to fairness, specifically highlighting and demonstrating mathematically that equity and equality are two somewhat contradictory central notions related to fairness in education. The contradictions revealed by them raise questions about the disparities between how AIED's diverse users are treated vs. how they are impacted by algorithmic interventions. While equality may be achieved through innovation if all individuals benefit the same amount regardless of their pre-existing capabilities, to achieve equity (e.g., closing the achievement gaps between learners from different socio-economic backgrounds), the impact of innovation must be positively greater for those with lower outcomes. This positioning presents a set of questions, likely some dilemmas, and further obligation of transparency for AIED designers with respect to both the form of algorithmic fairness they choose to furnish their systems with and the claims they can make about the generalisability of their applications to diverse users and contexts of use. When used as a research methodology AI can offer a means for systematic experimentation and interrogation in this context.

## Concluding remarks

Every domain of AI's potential application reflects the overarching ethical concerns discussed across the Ethical AI literature (e.g., Floridi and Cowles, 2019). However, each domain also brings very specific challenges that require not only close and contextualised examinations, but also actionable 'so whats' that can lead us out of the inertia fuelled by the rhetoric of the futility of any resistance against 'AI happening to us' into active explorations of the possible answers that illuminate, confront, and allow us to change the status quo. Equally, we do not want to resist for the sake of resistance. Instead, we are after a systematic and considered understanding of the risks as well as benefits of AI that can be used as a basis for informed debate, decision-making and refinement of our practices.

Throughout this book, there is a consistency with respect to the kinds of ethical challenges that the field of AIED faces and a set of proposals from across different perspectives for how we need to go-about questioning what AIED is, what role it plays and might need to play in society, and for how it can become a more defined and accountable discipline. There is a concerted call for questioning the assumptions on which AIED is based and for continuous involvement of multiple, diverse stakeholders in the design and auditing of AIED systems as well as the research methods and outcomes thereof that are employed within the field. No contribution in this book offers definitive answers. Instead, each chapter poses a series of critical questions for the AIED field. Sometimes these questions pertain to the kind of educational tools that AIED produces and their role in amplifying or ameliorating social inequities. In other instances, the questions focus on the ethical value of the assumptions we make in both how we design AIED systems and why, whereas in other cases, the authors pose fundamental philosophical questions about the role that AIED plays in shaping the way we educate the new generations. The totality of the contributions presented in this book, leads to a picture of AI in Education as a multifaceted discipline (as illustrated in **Figure 1**) which can act as a form of applied philosophy, a methodology for studying questions about learning and pedagogy, and as a form of civil engineering concerned with addressing immediate education-related challenges within society at large.

In the Introduction to this book, we suggested that a wider foundational perspective may need to be adopted by the AIED practitioners to engage actively in establishing the purpose of the AIED field (Holmes, Anastopoulou, et al., 2018; Kay, 2012). As has been illustrated through all of the contributions herein, the community is ready to step-up its engagement with the broader societal contexts which it has always claimed to serve to ensure that it is both *doing things ethically* and that it is *doing ethical things* (Holmes et al., 2021), and that its research and practices are able to stand to scrutiny within those wider contexts.

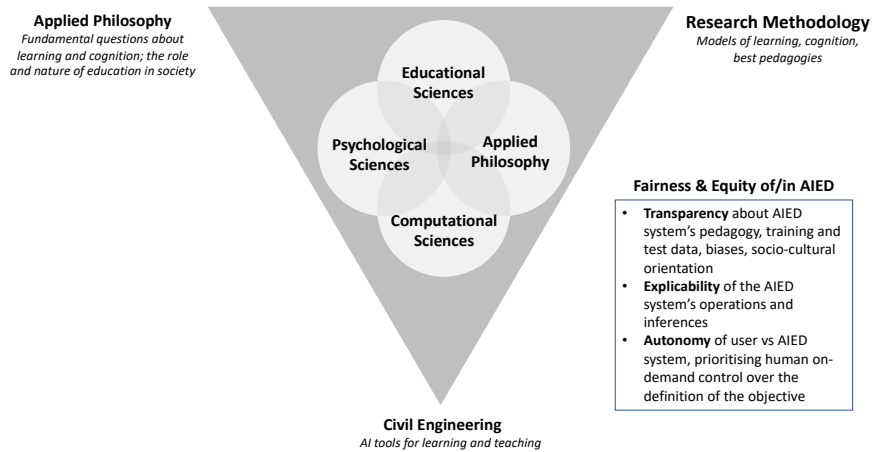

*Figure 1: The multifaceted nature of the AIED discipline*

If, as AIED practitioners often claim, such a grand ambition is to help make education and access to it fair and equitable, then a set of conditions, derived from the discussions throughout this book, need to be met (documented in the bottom right corner of Figure 1). First individual AIED systems need to be built for **transparency** to allow for inspection by different stakeholders of a multitude of assumptions on which they are based, including pedagogical assumptions (i.e., the pedagogical approaches that they encode), data on which their models have been constructed, and biases both in terms of their representational quality, including a declaration of the exact intended users, and the exact socio-cultural orientation of their curricular, pedagogical and communication models.

Second, the systems need to be built for **explicability** with respect to their operations (for example, explaining how the algorithms that underpin their models make inferences, what objective function or measure of success they follow, and what are their known inferential blind-spots, e.g., the generalisability across different students in different learning scenarios; levels of analytic granularity, etc.).

Third, the systems need to be built for human **autonomy**, giving different users the ability to modify, or to stop the operation of the system. How this is achieved exactly will be determined by who the user is and indeed what standard model remains at the base of such systems. In this respect, AIED stands at the forefront of developments in the broader AI, offering examples of AI-enabled approaches such as the open learner models and the teachable agents that challenge the standard model, at least of education, and where users are given different levels of control over their data, and the systems' inferences. This last dimension relates directly to Russell's call for the need to challenge the standard model on which our AI designs are based (discussed earlier), by removing the assumption of a perfectly known objective from the design of AI systems, and instead relying on the interaction between the human and the AI to negotiate the objectives for the individual users, with AI taking actions that expand rather than dictate human choice.

As such, we are optimistic that AIED can serve to examine grand questions about the meaning and purpose of education and about its own role in helping shape that purpose. Through the process of defining such a purpose AIED can help formulate research questions about human cognition and development, e.g., to spotlight the strengths and weaknesses of particular pedagogical supports in specific educational settings with specific types of learners, it can help address practical questions about what tools we can design, to whom they are targeted (to students directly, or to teachers to help them teach more effectively), and how we can deploy them to help our (transdisciplinary and multi stakeholder) grand ambitions come to fruition. With a mindset of ethical AIED, we are also optimistic about the community's ability and willingness to offer knowledge (rather than opinion) about where true opportunities for AI enhancement or change of human educational practices exist and provide scientific evidence of where caution or even opposition are needed.

Legg, S and Hutter M. (2007). Universal Intelligence: A Definition of Machine Intelligence. Minds & Machines 17, 391–444 (2007). https://doi.org/10.1007/s11023-007-9079-x.

Russell, S and Norvig, P. (1995). Artificial Intelligence: a modern approach. Prentice Hall

Russell, S. (2019). Human Compatible: Artificial Intelligence and the Problem of Control, Penguin, ISBN: 9780525558620

Vallor, S. (2021). The Thoughts The Civilized Keep. Noēma. https://www.noemamag.com/the-thoughts-the-civilized-keep

Weiner, N. (1960). Some Moral and Technical Consequences of Automation. Science,131,1355-1358,1960. American Association for the Advancement of Science.
## List of chapters in this book:

*Introduction* by Wayne Holmes and Kaśka Porayska-Pomsta

**Chapter 1:** Learning to learn differently by Jutta Treviranus

**Chapter 2:** Educational research and Artificial Intelligence in education: Identifying ethical challenges by Alison Fox

**Chapter 3:** AI in education: An opportunity riddled with challenges by Ivana Bartoletti

**Chapter 4:** Student-centered requirements for the ethics of AI in education by Lionel Brossi, Ana María Castillo and Sandra Cortesi

**Chapter 5:** Pitfalls and pathways for trustworthy Artificial Intelligence in education by Nathalie A. Smuha

*Chapter 6:* Equity and Artificial Intelligence in education: Will "AIED" *amplify* or *alleviate* inequities in education? by Kenneth Holstein and Shayan Doroudi

**Chapter 7:** Algorithmic fairness in education by René F. Kizilcec and Hansol Lee

*Chapter 8:* Beyond "fairness:" Structural (in)justice lenses on AI for education by Michael Madaio, Su Lin Blodgett, Elijah Mayfield, Ezekiel Dixon-Román

**Chapter 9:** The overlapping ethical imperatives of human teachers and their Artificially Intelligent assistants by Benedict du Boulay

**Chapter 10:** Integrating AI ethics across the computing curriculum by Iris Howley, Darakhshan Mir, Evan Peck

***Conclusions: Toward Ethical AIED*** by Kaśka Porayska-Pomsta and Wayne Holmes